\documentclass[prl,twocolumn,superscriptaddress,
showpacs,groupeaddress,preprintnumbers,amsmath,amssymb,tightenlines]{revtex4}
\usepackage{graphicx}
\newcommand{\beq}{\begin{equation}}
\newcommand{\eeq}{\end{equation}}
\newcommand{\bea}{\begin{eqnarray}}
\newcommand{\eea}{\end{eqnarray}}
\newcommand{\ba}{\begin{array}}
\newcommand{\ea}{\end{array}}
\newcommand{\bc}{\begin{center}}
\newcommand{\ec}{\end{center}}

\newcommand{\bml}{\begin{subequations}}
\newcommand{\eml}{\end{subequations}}
\newcommand{\commentout}[1]{{}}
\newcommand{\bk}{{\bf k}}

\newcommand{\K}{{\cal K}}
\newcommand{\adag}{a^\dagger}
\newcommand{\alphadag}{\alpha^\dagger}
\newcommand{\bdag}{b^\dagger}
\newcommand{\betadag}{\beta^\dagger}

\newcommand{\half}{\hbox{$\frac{1}{2}$}}

\newcommand{\HC}{{\rm H.c.}}
\newcommand{\eq}[1]{(\ref{#1})}
\newcommand{\etal} {{\it et al.\/}}
\newcommand{\ibid} {{\it ibid. \/}}
\newcommand{\vol}[1]{{\bf #1}}
\newcommand{\comment}[1]{{}}

\begin{document}
\title
{
Feshbach-Resonant Interactions in $^{40}$K and $^6$Li Degenerate Fermi Gases
}
\author{Matt Mackie}
\affiliation{QUANTOP--Danish National Research Foundation
Center for Quantum Optics, Department of Physics and Astronomy,
University of Aarhus, DK-8000 Aarhus C, Denmark}
\author{Jyrki Piilo}
\affiliation{School of Pure and Applied Physics,
University of KwaZulu-Natal, Durban 4041, South Africa}
\date{\today}

\begin{abstract}
We theoretically examine a system of Fermi degenerate atoms coupled to bosonic
molecules by a Feshbach resonance, focusing on the superfluid transition to a molecular
Bose-Einstein condensate dressed by Cooper pairs of atoms. This problem raises an
interest because it is unclear at present whether bimodal density distributions
observed recently in $^{40}$K and $^6$Li are due to a condensate of bosonic
molecules or fermionic atom pairs. As opposed to $^{40}$K,
we find that any measurable fraction of above-threshold bosonic molecules are
necessarily absent for the $^6$Li system in question, which strongly implicates
Cooper pairs as the culprit behind its bimodal distributions.
\end{abstract}


\maketitle

{\em Introduction.}--Magnetoassociation creates a molecule from a pair of
colliding atoms when one of the atoms spin flips in the presence of a magnetic field
tuned near a Feshbach resonance \cite{STW76}. Recently,
ultracold \cite{REG03} and condensate~\cite{GRE03} molecules have
been created via magnetoassociation of a Fermi gas of atoms, in the course of efforts to
create superfluid Cooper-paired atoms~\cite{REG04,ZWI04} (see also
Refs. \cite{CHI04}). In particular, for a magnetic field tuned above the
two-body threshold for molecule formation, the observation of a bimodal density
distribution for a system of $^{40}$K atoms was attributed to the existence of a
Bose-Einstein condensate of fermionic Cooper pairs~\cite{REG04}. Nevertheless, a lone
theoretical analysis suggests that the
$^{40}$K data\cite{REG04} can be understood as a Bose-Einstein
condensate of molecules, since the presence of the Fermi sea shifts the
threshold for molecular formation to the point where the molecular binding energy is
equal to twice the Fermi energy \cite{FAL04}, an interpretation that has been bolstered,
though not confirmed, by observations of bimodal distributions in $^6$Li
atoms \cite{ZWI04}. The purpose of this Letter is to demonstrate that significant
fractions of above-threshold molecular condensate are absent only when the atom-molecule
coupling is much larger than the Fermi energy. In a surprise reversal, our results point
to interpretations of a molecular and fermionic Bose-Einstein condensates, respectively,
instead of fermionic \cite{REG04} and molecular \cite{ZWI04} condensate.

\begin{figure}[b]
\centering
\includegraphics[width=8.0cm]{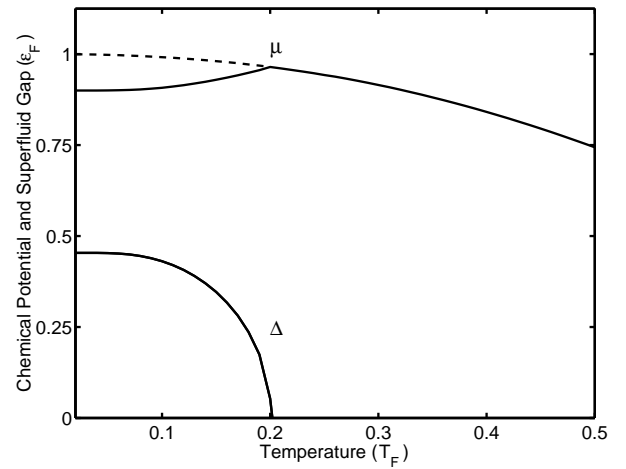}
\caption{Example of the onset of the superfluid transition to a Bose-Einstein condensate
of molecules dressed by fermionic atom pairs. The appearance of a non-zero superfluid gap
($\Delta$) lowers the chemical potential ($\mu$)
compared to the normal state (dashed curve), signaling the onset of the superfluid
regime. As it happens, the coupling constant and the detuning in this example are chosen
to be exceptionally large (see later text), similar to recent experiments~\cite{ZWI04}.}
\label{CUSP}
\end{figure}

The mean-field theory of magnetoassociation of a Fermi gas of atoms leads to two
types of instabilities against molecule formation. First is a dynamical instability,
whereby the larger state space of the molecules, owing somewhat to Pauli blocking,
leaves the atoms prone to spontaneous association into molecules~\cite{JAV04}. Here we
focus on the thermodynamic instability of a Fermi sea against the formation of Cooper
pairs~\cite{TIN75}, a trait of superconductors whose analog is passed on to
Feshbach-resonant superfluids~\cite{TIM01}. A thermodynamical instability
occurs because pairing lowers the energy, similar to Fig.~\ref{CUSP}, and so coupling to
a reservoir with a low enough temperature leaves the system prone to pairing. The
question is what role molecules play in this process.

Our answer is outlined as follows. After introducing the model~\cite{FRI89}, we show
that weak (strong) coupling gives a large (negligible) fraction of above threshold
molecules, and that, although contrary to two-body physics, this result
makes perfect sense in terms of our previous~\cite{JAV99,MAC02} boson results. In
particular, for atom-molecule couplings much larger than the Fermi energy, dissociation
to fermionic pairs should dominate the creation of bosonic molecules.
We then find that bosonic molecules can be ruled out for the observations
in Ref.~\cite{ZWI04}, but not for those in Ref.~\cite{REG04}. Before concluding we
contrast our results with the recent related
theories~\cite{FAL04,WIL04,DIE04,MEE04,AVD04}.

{\em Ideal Gas Model.}--We model an ideal two-component gas of fermionic
atoms coupled by a Feshbach resonance to bosonic molecules. An ideal
gas is chosen for simplicity, and is justified by a collisional interactions strength
that is negligible compared to the atom-molecule coupling (see next-to-last section).
In the language of second-quantization, an atom of mass $m$ and momentum
$\hbar\bk$ is described by the annihilation operator $a_{\bk,1(2)}$, and a molecule of
mass $2m$ and similar momentum is described by the annihilation operator $b_\bk$. All
operators obey their (anti)commutation relations. The microscopic Hamiltonian for such a
freely-ideal system is written:
\bea
\frac{H}{\hbar} &=&
\sum_{\bk} \left[\left(\epsilon_k-\mu\right)
  \adag_{\bk,\sigma}a_{\bk,\sigma}+(\half\epsilon_k+\delta_0-\mu_{\rm mol})\bdag_\bk
b_\bk\right]
\nonumber\\&&
-\frac{\K}{\sqrt{V}}
  \sum_{\bk,\bk'}\left(\bdag_{\bk+\bk'}a_{\bk,1}a_{\bk',2}+\HC\right),
\label{MICRO_HAM}
\eea
where repeated greek indices imply a summation ($\sigma=1,2$). The free-particle energy
is $\hbar\epsilon_{k}=\hbar^2 k^2/2m$, the atom (molecule) chemical potential is
$\hbar\mu_{\sigma(\rm mol)}$, and the bare detuning $\delta_0$ is a measure of the
binding energy of the molecule ($\delta_0>0$ is taken as above threshold), the
mode-independent atom-molecule coupling is
$\K$, and $V$ is the quantization volume. We have already imposed the ideal conditions
for superfluidity with $\mu_1=\mu_2=\mu$, and now we impose chemical equilibrium between
the atoms and molecules with $\mu_{\rm mol}=\mu_1+\mu_2=2\mu$. Diagonalization of the
Hamiltonian~\eq{MICRO_HAM} is achieved by the standard transformation to a dressed
basis~\cite{FRI89}:
\bml
\beq
\left(\begin{array}{c}\alpha_{\bk,1}\\ \alphadag_{-\bk,2}\end{array}\right)
  =\left(\begin{array}{cc}
    \cos\theta_k & -e^{i\varphi}\sin\theta_k \\
    e^{-i\varphi}\sin\theta_k & \cos\theta_k
  \end{array}\right)\!\!
  \left(\begin{array}{c} a_{\bk,1}\\ \adag_{-\bk,2}\end{array}\right)\!\!,
\label{BOGOa}
\eeq
\beq
\beta_\bk = b_\bk -\sqrt{V}\Phi\delta_{\bk,0},
\label{BOGOb}
\eeq
\label{BOGOT}
\eml
where $\delta_{\bk,0}$ is the Kronecker delta-function; hence,
\bea
\frac{H}{\hbar} &=& \left(\delta_0 -2\mu\right)V|\Phi|^2 
  +\sum_\bk
    \left(\half\epsilon_k+\delta_0-2\mu\right)\betadag_\bk\beta_\bk
\nonumber\\&&
  +\sum_\bk \left[ \left( \epsilon_k -\mu\right) 
    +\omega_k\left(\alphadag_{\bk,1}\alpha_{\bk,1}
      +\alphadag_{\bk,2}\alpha_{\bk,2}-1\right)\right].
\nonumber\\
\label{DIAG_HAM}
\eea
The condensate mean-field is
$\langle b_0\rangle/\sqrt{V}=e^{i\varphi}|\Phi|$, the mixing angle is
$\tan2\theta_k=|\Delta|/(\epsilon_k-\mu)$, the eigenfrequencies are
$\omega_k^2=(\epsilon_k -\mu)^2 +|\Delta|^2$, and the gap
is $|\Delta|=|\Phi|\K$.

To determine the thermodynamic ground state, we first calculate the pressure from the
partition function $\Xi=\text{Tr}\exp\left( -\beta H\right)$, which is then extremized
with respect to the molecular amplitude and, in turn, the chemical potential,
yielding~\cite{FRI89}:
\bml
\beq
(\delta-2\mu) = \Sigma(0)+
  \frac{\K^2}{2V}\,\sum_\bk\,
    \frac{1}{\omega_k}\tanh\half\beta\omega_k.
\label{CHEMPOT}
\eeq
\bea
\rho &=& 2|\Phi|^2+\frac{2}{V}\sum_\bk
  \frac{1}{\exp\left[\beta\left(\half\epsilon_k+\delta-2\mu\right)\right]-1}
\nonumber\\&&
  +\frac{1}{V}\sum_\bk\frac{\omega_k +\mu -\epsilon_k 
    +\left( \omega_k -\mu +\epsilon_k\right)\exp(-\beta\omega_k)}
{\omega_k\left[1+\exp(-\beta\omega_k)\right]}\,.
\nonumber\\
\label{DENS}
\eea
\label{ALGEBRAIC}
\eml
Renormalization is via the resonant self-energy $\Sigma(0)$~\cite{MAC02}, meaning the
physical detuning $\delta$ replaces the
bare $\delta_0$.

\begin{figure}
\centering
\includegraphics[width=8.0cm]{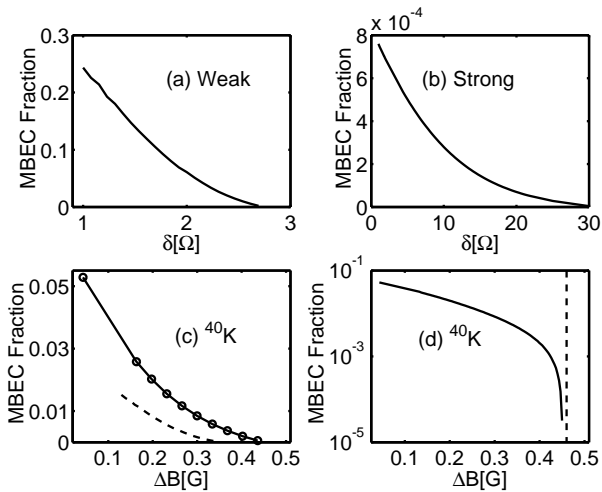}
\caption{Molecular condensate fraction as a function of the above-threshold
detuning. For
$T=0.05T_F<T_C$, where $T_C$ is the critical temperature for the superfluid
transition, weak coupling ($\Omega=\epsilon_F$) finds a large
fraction of above-threshold molecular condensate~(a), whereas a strong coupling
($\Omega= 35\epsilon_F$) finds a largely absent MBEC~(b), independent of the
above-threshold detuning. In panel~(c), the solid line is for $T=0.08T_F<T_C$
and $\Omega=4\epsilon_F$, and the dashed line is the weak result from panel~(a)
scaled by $1/4^2$~\cite{SCALE}. The open circles are used to emphasize how only the first
point is taken from the region where the ideal gas approximation may break down. The
actual results are markedly better than the scaling, and in agreement with
observation~\cite{REG04} except for a shift of about 0.2~G. In panel~(d) the superfluid
$^{40}$K phase boundary is evident in a departure from exponential behavior.}
\label{MBEC_FRAC}
\end{figure}

{\em Weak vs. Strong Coupling.}--Solving the algebraic system~\eq{ALGEBRAIC}
self-consistently determines the chemical potential as a function of temperature.
Intuitively the appearance of a non-zero gap lowers the chemical potential of the
superfluid BEC-pair dressed state compared to the normal state, which allows us to
confirm that, for a given temperature, the system is in the superfluid regime, c.f.,
Fig.~\ref{CUSP}. Along with the conservation of particle number,
Figs.~\ref{MBEC_FRAC}~(a,b) indicates a ground state that is a molecular
condensate dressed by dissociated fermionic pairs, an admixture
adjustable according to the detuning from threshold. In particular, for an atom-molecule
coupling that is weak compared to the Fermi energy,
$\Omega=\sqrt{\rho}\K\alt\epsilon_F$ [where $\epsilon_F=\hbar(3\pi^2\rho)^{2/3}/2m$], we
find a large fraction of above-threshold molecular condensate [Fig.~\ref{MBEC_FRAC}~(a)],
whereas $\Omega\gg\epsilon_F$ leads to an above-resonance system that is
predominantly fermionic pairs [Fig.~\ref{MBEC_FRAC}~(b)].

We have confirmed that for strong coupling the
fraction of molecular condensate remains negligible ($\sim10^{-3}$) for
$\delta\sim\epsilon_F$, i.e., the absence of a large near-threshold fraction is not the
result of choosing $\Omega$ as the frequency scale; similarly, having properly
renormalized the detuning, it is not due to any spurious shift of resonance threshold.
Additionally, the trap, albeit omitted, can actually favor the occurrence of superfluid
pairing~\cite{SEA02}.

Below threshold ($\delta<0$), Fourier analysis delivers the binding
energy, $\hbar\omega_B<0$, of the Bose-condensed molecules~\cite{JAV04,MAC02}:
$\omega_B-\delta-\Sigma'(\omega_B)+i\eta=0$, where $\Sigma'(\omega_B)$ is the finite
self-energy of the Bose molecules and
$\eta=0^+$. Above the two-body threshold ($\delta>0$) gives an imaginary
$\omega_B$, and the bound state ceases to exist; nevertheless,
Fig.~\ref{MBEC_FRAC}~(a) shows a large fraction of molecular condensate, which drops off
for increasing coupling strength as per Fig.~\ref{MBEC_FRAC}~(b). These apparently
contradictory results are in fact consistent with the dynamical studies of rapid
adiabatic passage in bosons~\cite{MAC02,JAV99}: just above threshold, the coupled
atom-molecule system can have a significant fraction of molecular
condensate~\cite{JAV99}, but ``rogue" dissociation to atom pairs with equal and opposite
momentum is expected to dominate the formation of molecules for
$\Omega\gg\hbar\rho^{2/3}/m\approx\epsilon_F$~\cite{JAV99,MAC02}. Physically, the
rate for converting atoms into molecules is $\sim\Omega$, whereas the rate
for rogue magnetodissociation is $\Gamma_0\propto\Omega^2$~\cite{FOOTIE}; hence,
for a strong enough coupling, rogue magnetodissociation to fermionic pairs of
atoms with equal-and-opposite momentum will dominate the formation of molecular
condensate.

{\em Recent Experiments and Theory.}--The significance of the above results
is seen by comparison with experiments in $^{40}$K~\cite{REG04} and
$^6$Li~\cite{ZWI04} systems. Detunings are converted into magnetic fields according to
$\Delta B=\hbar\delta/\Delta_\mu$, where the difference in magnetic moments between the
atom pair and a molecule is $\Delta_\mu$, and where $\Delta B=B-B_0$ (with $B_0$ the
magnetic-field position of resonance).

For a $^{40}$K gas of density
$\rho=2\times10^{13}\rm{cm}^{-3}$, the coupling strength is~\cite{JAV04,HUL04}
$\Omega\approx4\epsilon_F$; the difference in magnetic moments is
$\Delta_\mu\approx0.19\mu_0$~\cite{JAV04} (with $\mu_0$ the Bohr magneton). Off hand, we
are tempted to scale~\cite{SCALE} the results of Fig.~\ref{MBEC_FRAC}~(a), but the
result is not encouraging [Fig.~\ref{MBEC_FRAC}~(c), dashed line]; nevertheless, a full
recalculation~[Fig.~\ref{MBEC_FRAC}~(c), solid line and
open circles] of the solution to Eqs.~\eq{ALGEBRAIC} for $\Omega=4\epsilon_F$ yields
results that, except for a roughly 0.2~G shift, agree embarrassingly well with the
measured~\cite{REG04} bimodal distributions (not shown). The 0.2~G disagreement is
understood more clearly given Fig.~\ref{MBEC_FRAC}~(d), where the superfluid phase
boundary in detuning, given roughly by the dashed line, is marked by a clear departure
from exponential behavior. Overall, it seems that we agree with
Ref.~\cite{FAL04} that  bosonic molecules can not be ruled out, by two-body physics or
otherwise, as the culprit responsible for the condensate footprints in
$^{40}$K~\cite{REG04}.

On the other hand, for a $^6$Li gas of typical density~\cite{ZWI04}, the atom-molecule
coupling is~\cite{HOU98,HUL04} $\Omega=87\epsilon_F$. Approximating
$\Delta_\mu=2\mu_0$, any molecular BEC can already be ruled out of
experiments~\cite{ZWI04} based on Fig.~\ref{MBEC_FRAC}~(b). Even without explicitly
accounting for the somewhat larger $^6$Li coupling constant, Fig.~\ref{MBEC_FRAC}~(b)
would predict a near-unit fraction of fermionic pairs. Since the atom-molecule coupling
in Fig.~\ref{CUSP} is $\Omega\sim150\epsilon_F$ and the detuning is $\delta\gg\Omega$,
we estimate the critical temperature $T_C\sim0.2T_F$, which is close to the
measured~\cite{ZWI04} large-detuning critical temperature (all things considered);
hence, we expect $T=0.05T_F$ to be far enough below the critical temperature so that
what is not molecular condensate is most likely superfluid fermionic pairs, with only a
small thermal fraction. Indeed, whereas initial experiments
measure 80\% condensate fractions~\cite{ZWI04}, recent measurements are in excess of
90\%~\cite{ZWI04b}, consistent with expectation.

We pause to briefly justify the ideal gas model. The collisional interaction strength
is $\Lambda=2\pi\hbar\rho a/m$, where $a$ is the off-resonant atomic $s$-wave scattering
length. The $^{40}$K and $^6$Li scattering length are
roughly an order of magnitude apart: $a_{\rm K}=176a_0$~\cite{BOH00} and
$|a_{\rm Li}|=2110\,a_0$~\cite{HOU98,BAR04}, with $a_0$ the Bohr radius. For a
typical density $\rho\sim10^{13}\rm{cm}^{-3}$, it turns out that magnitude of the
collisional coupling strengths, in units of the atom-molecule coupling, are roughly
equal: $|\Lambda|/\Omega\approx10^{-3}$. Collisions should therefore be broadly
negligible.

Before closing, it is important to draw contrast with the latest work of
others. First, it appears that the stability of the above-threshold
molecules is a matter of competition between formation of molecular
condensate and the subsequent magnetodissociation to fermionic pairs, as opposed to
many-body effects~\cite{FAL04} (see also Ref.~\cite{WIL04}). Next, although the chemical
potential for the near-resonant $^{40}$K system is undoubtedly in the so-called
universal regime~\cite{DIE04}, and molecules are not expected to play role in such a
case~\cite{DIE04}, we found herein that molecules cannot be ruled out. Also, whereas a
prominent single-channel theory~\cite{KOK02}--i.e., only atoms and their pairs, no
explicit molecules--has shown good agreement with the $^{40}$K experiments~\cite{AVD04},
this theory would presumably deliver similar answers for both
experiments~\cite{REG04,ZWI04}, unlike our theory. Finally, it is entirely possible that
the threshold of a strongly-coupled system is systematically shifted to negative
detunings, as suggested recently~\cite{MEE04}--albeit without recourse to the
atom-molecule coupling strength, a matter defered to elsewhen.

{\em Summary and Conclusions.}--We have found that, in a Feshbach-resonant gas of Fermi
atoms, significant fractions of molecular condensate are absent for atom-molecule
couplings that are strong compared to the Fermi energy. While it is perhaps a stretch
call the $^{40}$K~\cite{REG04} system weak, it is clear that bosonic molecules can not be
ruled out of its bimodal distributions. However, bosonic molecules can be ruled out for
the $^6$Li~\cite{ZWI04} system, which strongly suggests that Cooper
pairing/fermionic condensation has been observed. Our interpretation is that, because
rogue magnetodissociation favors fermionic pair formation, a condensate of Cooper pairs
rather than molecules is formed. These results suggest that, in addition to bosons and
fermions, the schism between dynamics and thermodynamics is, at least on some level,
artificial. Moreover, debate is currently raging over the necessity of a separate
molecular channel in theories of Feshbach-resonant fermionic atoms~\cite{KITP}, and our
results strongly suggest that molecules play an explicit role in a wholistic
understanding of experiments.

{\em Acknowledgements.}--We kindly thank Michael Budde, Eric Cornell, Jason Ho, Randy
Hulet, Nicolai Nygaard, Henk Stoof, and Martin Zwierlein for valuable insight, as well as
the Academy of Finland (JP, project 204777) and the Magnus Ehrnrooth Foundation for
support (JP).


\begin{thebibliography}{99}
\bibitem{FOOTIE}
  The simplest way to infer these rates is by the Heisenberg equations of
motion for the Hamiltonian~\eq{MICRO_HAM}.
\bibitem{SCALE}
  Based loosely on Timmermans \etal~\cite{TIM01}, we find that the molecular fraction
    $f=2|\Phi|^2/\rho$ scales like
    $f/f_0=(\bar\Omega_0/\bar\Omega)^2$, where
    $\bar\Omega=\Omega/\epsilon_F$.
\end{thebibliography}
\end{document}